\documentclass[twocolumn]{aastex63}

\newcommand{\lya}{Ly$\alpha$}
\newcommand{\Lya}{Ly$\alpha$\space}
\newcommand{\NHI}{$N_{\rm HI}$}
\newcommand{\vexp}{$v_{\rm exp}$}
\newcommand{\fesc}{f_{\rm esc}}
\newcommand{\dd}{\mathrm{d}}
\newcommand{\kms}{\,\ifmmode{\mathrm{km}\,\mathrm{s}^{-1}}\else km\,s${}^{-1}$\fi\space}
\newcommand{\cm}{\,\ifmmode{\mathrm{cm}}\else cm\fi\space}

\submitjournal{ApJ}

\shorttitle{Can galaxy evolution mimic cosmic reionization?}
\shortauthors{Hassan \& Gronke}
\graphicspath{{./}{figures/}}
\begin{document}

\title{Can galaxy evolution mimic cosmic reionization?}

\correspondingauthor{Sultan Hassan}
\email{shassan@flatironinstitute.org}

\author[0000-0002-1050-7572]{Sultan Hassan}
\altaffiliation{Flatiron fellow}
\affiliation{Center for Computational Astrophysics, Flatiron Institute, 162 5th Ave, New York, NY 10010, USA}\affiliation{Department of Astronomy, New Mexico State University, Las Cruces, NM 88003, USA}\affiliation{Department of Physics \& Astronomy, University of the Western Cape, Cape Town 7535,
South Africa}

\author[0000-0003-2491-060X]{Max Gronke}
\altaffiliation{Hubble fellow}
\affiliation{Department of Physics \& Astronomy, Johns Hopkins University, Baltimore, MD 21218, USA}

\begin{abstract}
Lyman-$\alpha$ (Ly$\alpha$) emitting galaxies are powerful tools to
probe the late stages of cosmic reionization. The observed sudden drop
in Ly$\alpha$ fraction at $z>6$ is often interpreted as a sign of
reionization, since the intergalactic medium (IGM) is more neutral and opaque to Ly$\alpha$ photons.
Crucially, this interpretation of the
observations is only valid under the assumption that galaxies
themselves experience a minimal evolution at these epochs. By
modelling Ly$\alpha$ radiative transfer effects in and around
galaxies, we examine whether a change in the galactic properties can
reproduce the observed drop in the Ly$\alpha$ fraction. We find that
an increase in the galactic neutral hydrogen content or a reduction in
the outflow velocity toward higher redshift both lead to a lower
Ly$\alpha$ escape fraction, and can thus mimic an increasing neutral
fraction of the IGM. We furthermore find that this change in galactic
properties leads to systematically different Ly$\alpha$ spectra which can be
used to differentiate the two competing effects. 
Using the CANDELSz7 survey measurements which indicate slightly broader lines at $z\sim 6$,  we find that the scenario of a mere increase in the galactic column density towards higher $z$ is highly unlikely. We also show that a decrease in outflow velocity is not ruled out by existing data but leads to more prominent blue peaks at $z>6$.
Our results caution
using Ly$\alpha$ observations to estimate the IGM neutral fraction
without accounting for the potential change in the galactic properties, e.g., by  mapping out the evolution of Ly$\alpha$ spectral characteristics.

\end{abstract}



\section{Introduction} \label{sec:intro}
Lyman-$\alpha$ (\lya) line is a promising tool to probe cosmic reionization as the increasingly neutral intergalactic medium (IGM) becomes more opaque to \Lya photons towards higher redshifts \citep[e.g. for extensive review see][]{Dijkstra:2014}. This increased optical depth is expected to give rise to a decrease in the observed number of \Lya emitting galaxies at $z\gtrsim 6$. Specifically, the number of \Lya selected galaxies (or Lyman-$\alpha$ emitters, LAEs) decreases dramatically at this redshift \citep[e.g.,][]{2010ApJ...722..803O,2010ApJ...721.1853T,2014MNRAS.440.2375M,2016MNRAS.463.1678S,2018PASJ...70S..16K}. Similarly,  continuum selected, or Lyman break galaxies (LBGs) show a modest increase of \lya\ emission from $z\sim3$ to $z\sim6$ \citep[e.g.,][]{2011ApJ...730....8H,2018MNRAS.476.4725S}, and sudden drop at $z\geq6$ \citep{Barros:2017,Caruana:2014,Kusakabe:2020,Mason:2018,Mason:2019,Pentericci:2014,Pentericci:2018,Schenker:2014,Stark:2011,Tilvi:2014,Treu:2013,2019ApJ...878...12H,Jung2020}.
Especially the latter is a powerful observational probe -- as the Lyman break techniques allows the efficient detection of high-$z$ galaxies -- and is commonly parametrized by the `\Lya fraction' which describes the fraction of LBGs possessing a \Lya equivalent width $W > W_{\rm c}$ where $W_{\rm c}$ is an observationally determined cutoff, usually $20\,$\AA.

These different \Lya based observations are being used to constrain the evolution of the cosmic neutral fraction \citep{2006MNRAS.365.1012F,2007MNRAS.381...75M,2016MNRAS.463.4019K,Mason:2018,Mason:2019,2020ApJ...892..109N,2020MNRAS.495.3602W}.
In fact, at $z\sim 7$ these measurements pin the neutral fraction to, for instance, $\langle x_{\rm HI}\rangle \approx 0.59_{-0.15}^{+0.11}$ ($1\sigma$ error; taken from \citealp{Mason:2018}; other studies such as the ones mentioned above use a different set of assumptions and, thus, conclude a different evolution of $x_{\rm HI}$) and are, thus, currently more constraining than measures of the cosmic microwave background \citep{PlanckEoR} or quasar proximity zones \citep[e.g.,][]{Greig2017MNRAS.466.4239G,Davies2018ApJ...864..142D,2020ApJ...897L..14Y,2020ApJ...896...23W}.

However, these constraints are crucially dependent on the assumption that the average \textit{galactic} \Lya escape fraction does not change over this redshift interval as the observed \Lya flux is proportional to this times the intergalactic transmission. Therefore, an evolution in the cosmic neutral fraction is fully degenerate with the evolution of the \Lya escape fraction. 
While on the one hand, the duration from $z=7$ to $z=6$ is merely $\sim 170$ million years, i.e., relatively short in terms of galaxy evolution, one should keep in mind that this notion stems from of studies carried out at lower redshifts. % interstellar and circumgalactic medium at lower redshift

Importantly, \Lya is a resonant line with a large cross section which implies that \Lya escape through the interstellar and circumgalactic medium is a highly non-linear process. Several theoretical studies have shown that \Lya escape is dependent not only on the dust and neutral hydrogen abundance \citep{Neufeld1990,Dijkstra2006} but also on its kinematics \citep{Bonilha1979,Zheng2013}, and structure \citep{Neufeld1991,Gronke2017}, and that even small changes in these properties can have large effects on the \Lya observables -- and, in particular, the escape fraction.

Independently of the question whether the currently employed assumption of a constant \Lya escape fraction with redshift is justified, it is important to incorporate our ignorance regarding the evolution of the interstellar and circumgalactic medium into the models constraining cosmic evolution \citep[cf. work by][suggesting a larger accretion rate leading to a larger gas reservoir, and thus, lower \Lya escape fraction]{2011MNRAS.412.1123P,2012ApJ...756..164F}. 
\citet{Sadoun:2017} took a first strive at this goal by demonstrating that the observed drop in \Lya fraction can be entirely due to the increased neutral hydrogen content in the infalling region surrounding the dark matter halo hosting the galaxy. 
While in their interpretation this increased neutral fraction is due to a change in the ionzing background -- and, thus, arguably also a sign of cosmic reionization -- this result is very important as it shows the potential impact of this change of \Lya transmission not stemming from an evolution of the intergalactic medium.

In this paper, we want to systematically explore what changes in galactic properties can mimic the observed evolution of \Lya visibility usually attributed to the Epoch of Reionization. We will, furthermore, study how such changes will impact the \Lya spectra. This will allow future studies to use this additional constraints, and thus allow them to fold in the uncertainty regarding the galactic evolution into the models. 

This paper is organized as follows: in Sec.~\ref{sec:methods}, we describe the quantities and the radiative transfer code used, in Sec.~\ref{sec:results} we present our results, and we discuss them in Sec.~\ref{sec:conclusion}.

\section{Methods}
\label{sec:methods}

\subsection{\lya\ fraction}
As stated above, the \lya\ fraction, $X_{\rm Ly\alpha}$, is commonly defined as the fractional abundance of galaxies with \lya\ equivalent width ($W$) above certain cut-off ($W_{\rm c}$), which can be written as:
\begin{equation}\label{eq:xfrac}
    X_{\rm Ly\alpha, W_{c}} = \int^{\infty}_{W_{c}} p(W) \,\dd W,
\end{equation}
where $p(W)$ is the equivalent width distribution function. As commonly used in the literature \citep{2012MNRAS.419.3181D,2015MNRAS.449.1284G,Sadoun:2017}, we adopt an exponential form for $p(W)$:
\begin{equation}
p(W) = \frac{\exp{ \left(-W/W_{0} \right)} }{W_{0}+W_{1}}\text{\qquad for } W>0,  
\end{equation}
where $W_{0}$ and $W_{1}$ are free parameters, which can be found by matching to observations. It has been found that this parametrization reproduces observations reasonably well \citep{2014ApJ...795...20S}\footnote{Note while we focus on the evolution in the \Lya fraction, a change in the observed equivalent widths, e.g., due to a change in galactic properties, $W\rightarrow a W$ leads to a change in the scale height $W_0 \rightarrow a W_0$. This means that our results can be understood as a change in the equivalent width distribution. The details depend naturally on the parametrization of the EW distribution.}.

By integrating Equation~\ref{eq:xfrac} using  two  different observationally-motivated thresholds $W_{\rm c} = 25, 55$ \AA, the exponential scale $W_{0}$ becomes:
\begin{equation}
 W_{0} = \frac{30\,\text{\AA}}{\ln(X_{{\rm Ly\alpha},25} / X_{{\rm Ly\alpha},55})}\;.
\end{equation}
The $W_{0}$ is $M_{\rm UV}$ magnitude dependent as observations indicate. Using the measured \lya\ fractions at $z=6$ by~\citet{Stark:2011}, we find $W_{0,z=6}= 43.3, 30.2$ \AA\, for the faint ($M_{\rm UV} > -20.25$) and bright ($M_{\rm UV} < -20.25$) populations, respectively. We will use this $W_{0,z=6}$ value for the faint population throughout. By doing so, we match the observed \lya\ fraction measurement at $z=6$. We then attempt to reproduce the observed drop in \lya\ fraction at $z>6$ by changing only the galactic properties. 

In general, the observed \lya\ equivalent width $W$ is given by
\begin{equation}
    W = \frac{f_{\rm esc, Ly\alpha}}{f_{\rm esc, UV} } T_{\rm IGM} W_{\rm i}\,,
\end{equation}
where $T_{\rm IGM}$ is the IGM transmission, $W_{\rm i}$ is the intrinsic equivalent width, and $f_{\rm esc, Ly\alpha}$ and $f_{\rm esc, UV}$ are the photon escape fractions for \lya\ and  UV photons respectively. We assume that the IGM doesn't evolve (i.e. $T_{\rm IGM}=\mathrm{const.}$,  see \S\ref{sec:conclusion} for more details), and the dust optical depth $\tau_{\rm d}$ is the same at $z\geq6$, which also keeps the $f_{\rm esc, UV}$ constant ($f_{\rm esc, UV} = \exp(-\tau_{\rm d})$).  With these assumptions, we can translate the change in the equivalent width distribution to change in the \lya\ photon escape fraction. By simultaneously solving the \lya\ fraction equations for $z=6$ and $z>6$, we obtain
\begin{equation}\label{eq:lya_fesc}
\frac{ f_{\rm esc, Ly\alpha} (z>6)  }{ f_{\rm esc, Ly\alpha} (z=6)  } = \left[1-  \frac{W_{0,z=6}}{W_{c}} \log\left(  \frac{X_{{\rm Ly\alpha},c}(z>6)}{X_{{\rm Ly\alpha},c}(z=6)} \right)  \right]^{-1}\, .
\end{equation}
This equation relates the change in the photon escape fraction to the change in the \lya\ fraction between two redshifts. This relation depends on the equivalent width cut-off $W_{\rm c}$ and the exponential scale $W_{0}$ for the equivalent width distribution, which is a magnitude dependent. 

Here we focus on the measurements for the faint populations ($M_{\rm UV} > -20.25$) with cut-off $W_{c} > 25$\AA\,.
Our results might be quantitatively different for the bright population, but nevertheless the qualitatively result will remain unchanged, which is main focus for the work is to test the scenarios with which the change in the galactic properties can mimic reionization sign. We will come to this point later in the discussion.

There are several measurements for the \lya\ fraction at $z\geq 6$. For simplicity, we take the average
values from~\citet{Stark:2011} and~\citet{Barros:2017} at $z\sim6$,~\citet{Pentericci:2014},~\citet{Schenker:2014},~\citet{Caruana:2014},~\citet{Mason:2018} at $z\sim7$,~\citet{Tilvi:2014},~\citet{Treu:2013},~\citet{Schenker:2014} and~\citet{Mason:2019} at $z\sim8$. These average values of the \lya\ fraction are 0.46, 0.24 0.14 at $z\sim 6,7,8$, respectively. Using Equation~\ref{eq:lya_fesc}, these average values indicates that the photon escape fraction may equivalently drop by $f_{\rm esc, Ly\alpha} (z=7) /f_{\rm esc, Ly\alpha} (z=6) = 0.47$, and $f_{\rm esc, Ly\alpha} (z=8) /f_{\rm esc, Ly\alpha} (z=6) = 0.33$, to mimic the observed drop in \lya\ fraction from $z=6$ to $z=7$ and $z=6$ to $z=8$, respectively. It is worth noting that if we use~\citet{Pentericci:2018} measurements at $z=6$, which are lower than those complied by~\citet{Stark:2011}, to find the exponential scale $W_{0}$, the required drop would be  $f_{\rm esc, Ly\alpha} (z=7) /f_{\rm esc, Ly\alpha} (z=6) = 0.53$, and $f_{\rm esc, Ly\alpha} (z=8) /f_{\rm esc, Ly\alpha} (z=6) = 0.39$. These values are still consistent within 1-$\sigma$ level of each other, and would not significantly alter our conclusion. Our aim is to study the conditions with which a change in the galactic properties leads to these differences in the photon escape fraction, and hence mimicking reionization. While uncertainties in the measurements do exist \citep[see, e.g.,][]{Kusakabe:2020,Jung2020}, the outcome of our theoretical study does not depend on the exact observationally inferred drop of the \Lya visibility. Instead, we want to explore here whether in principle such a drop can be reproduced by evolution of galactic properties.

\subsection{Monte-Carlo \lya\ Radiative Transfer}
We model \lya\ emission from and around galaxies assuming shell models as implemented within a Monte-Carlo radiative transfer (MCRT) code {\sc tlac}~\citep{Gronke:2014}. The MCRT methods tracks the evolution and properties of injected photons including direction and frequency as they travel through the simulation domain.
The `shell model' is commonly adopted as it has been shown to reproduce observed \Lya spectra well \citep{Ahn_2003,Verhamme2006A&A...460..397V,Gronke2017A&A...608A.139G}\footnote{There is an ongoing discussion in the literature regarding the physical meaning behind the `shell model' \citep[e.g.][]{Orlitova2018A&A...616A..60O,Gronke2017}. We will comment on the interpretation of our results in light of the adopted model in \S\ref{sec:conclusion}.}.
It is defined by the neutral hydrogen column density \NHI\,, the dust optical depth $\tau_{\rm d}$, the expanding / outflowing velocity \vexp, and the effective temperature $T$.
In all our runs, we consider an initial number of photon packages of $N=10^{5}$ which we inject at line center, unless otherwise stated. We here consider two scenarios to the change in the galactic properties that can lead to a change in the photon escape fraction required to mimic reionization. While keeping all other properties fixed, these scenarios are changing only either the column density \NHI\, or the outflows \vexp\,. We leave $T=10^4\,$K fixed for all the runs.

To this end, we have shown how the drop in \lya\ fraction is equivalent to a drop in the photon escape fraction while keeping the IGM fixed. In the next section, we present the our results relating the change in the galactic properties to the photon escape fraction and spectral properties, as well as comparing with observations to discriminate between these two scenarios.
\begin{figure*}[ht!]
    \centering
    \includegraphics[scale=0.5]{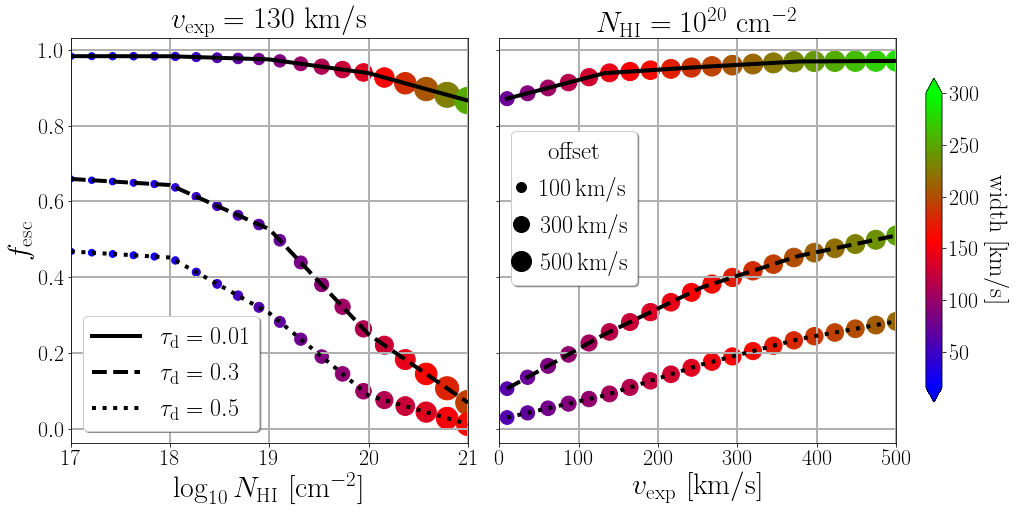}
    \caption{The \lya\ photon escape fraction $\fesc$ dependence on the galactic properties. {\it Left}: $\fesc$ as a function of column density \NHI\ at fixed outflow velocity \vexp$=130\, {\rm km/s}$.  {\it Right}: same as left but as a function of \vexp\ at a fixed \NHI$=10^{20} {\rm cm^{-2}}$\,. Different linestyles correspond to different dust amounts $\tau_{\rm d}$ as shown in the legend. Different colors and point sizes represent the spectral width (square root of second moment) of the red peak and offset of the red side of \lya\ emission, respectively. $\fesc$ decreases in denser and dustier media and increases with higher outflows. The width and offset both increase with increasing column density and outflow velocity.  An increase in the column density \NHI\ or decrease in outflows \vexp\ towards high redshift ($z>6$) by a factor of $\sim\times\, 2-3$ reproduces the observed drop in \lya\ fraction $X_{\rm Ly\alpha}$ and mimics the increase in neutral fraction (i.e. reionization) without an evolving IGM.} 
    \label{fig:fesc_NHI_VEL}
\end{figure*}

\section{Results}
\label{sec:results}

\subsection{Impact of galactic properties on $f_{\rm esc}$ and \Lya spectra}
Figure~\ref{fig:fesc_NHI_VEL} is a visual summary of how the \lya\ photon escape fraction $\fesc$ and spectral properties change as the galactic properties evolve.
We show the dust impact on $\fesc$ with variation in the optical depth $\tau_{\rm d}$ as quoted in the legend and represented by different linestyles. We color-code $\fesc$ dependence on the galactic properties (\NHI\ \& \vexp) with  the width (square root of second moment), and the point sizes reflect the offset (the first moment) of the red side of \lya\ emission.

The left panel of Fig.~\ref{fig:fesc_NHI_VEL} shows the $\fesc$ dependence on the column density \NHI\ at a fixed outflow velocity \vexp$=130\,{\rm km/s}$, whereas the right panel depicts the dependence on the outflow velocity \vexp\ at a fixed column density \NHI$=10^{20} {\rm cm^{-2}}$. In general, $\fesc$ decreases as the \NHI\ increases and \vexp\ decreases.
Both a higher HI column density and a lower outflow velocity imply that the optical depth at line-center increases, and thus, so does the path length of \lya\ photons through the scattering medium. This in turn means that the effective dust optical depth increases, lowering the escape fraction.

Note that the $\fesc$ dependence on \NHI\ is steeper than that on \vexp. At small dust amounts (solid lines), $\fesc$ is almost unity as most of photons easily escape. In this regime, the galactic properties are required to change dramatically in order to observe a factor of 2 difference in $\fesc$. With a dustier medium (dashed and dotted lines), it is easier to find such a difference with smaller change in the galactic properties.  For instance, at $\tau_{\rm d}\geq0.3$, a change by $\leq$ 1 order of magnitude in \NHI\ or by $\leq$ 200 $\rm km/s$ in \vexp\ is needed to reduce $\fesc$ by factor 2. 
We also see that the spectral properties such as the width and offset change accordingly. These changes can potentially be tested against observations (cf. \S\ref{sec:comparison_obs} below).
The width and offset both increase as the \NHI\ or/and \vexp\ increase. Similarly, the dependence on \NHI\ is steeper since we see the width change from about $\sim$ 50 $\rm km/s$ at $\log_{10}$ \NHI\ / $\rm cm^{-2}=$17 up to more than 500 $\rm km/s$ by $\log_{10}$ \NHI\ / $\rm cm^{-2}=$21.
Overall we find a tight correlation between offset and width, with the offset being roughly about $\sim 2$ times the width, in agreement with previous studies \citep[e.g.,][]{Neufeld1990,Verhamme2018MNRAS.478L..60V}.
Unsurprisingly, the spectral properties are mostly unaffected by the dust content.
On the other hand, the line width dependence on \vexp\ somewhat modest as the it changes from 300 $\rm {\rm km/s}$ at very low \vexp\ $\sim\, 5  {\rm km/s}$ to about less than 500 $\rm km/s$ by \vexp\ $\sim 500\, {\rm km/s}$. Likewise, the offset dependence on \NHI\ is more significant since point sizes change significantly towards high \NHI\ values as opposed to the slow change as the \vexp\ increases.

\begin{figure*}[ht!]
    \centering
    \includegraphics[width=.95\textwidth]{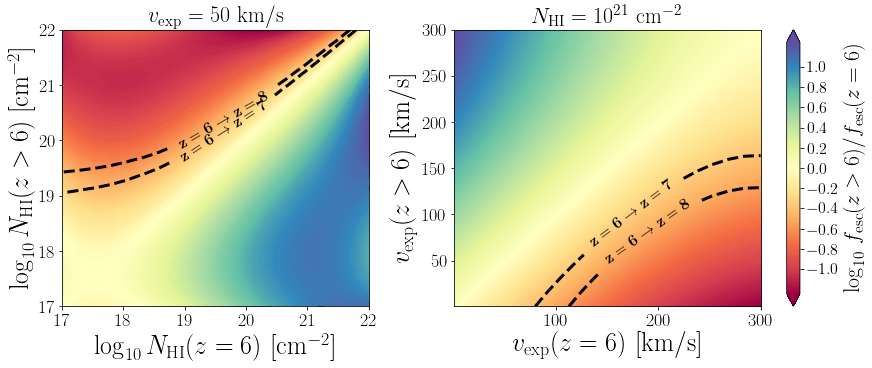}
    \caption{Grid of models with different column densities \NHI\ and fixed outflow velocity \vexp\ $=50\, {\rm km/s}$ (left) and with different outflows \vexp\ at fixed column density \NHI\ $=10^{21}\, {\rm cm^{-2}}$ (right), both at the same amount of dust $\tau_{\rm d}=0.35$  at $z\geq 6$. The horizontal axis represents the column density and outflows at $z=6$ whereas the vertical axis shows the same quantities at higher redshifts, $z=7,\, 8$. Both panels are color-coded with photon escape fraction $\fesc$ ratio between $z>6$ and $z=6$. Contours show the possible column densities/outflows between $z=6\rightarrow z=7$ and $z=6\rightarrow z=8$ with $\fesc$ difference that mimics the sign of reionization as often inferred from the observed drop in \lya\ fraction $X_{\rm Ly\alpha}$ at these epochs. }
    \label{fig:cont_NHI_VEL}
\end{figure*}

In summary, Fig.~\ref{fig:fesc_NHI_VEL} illustrates nicely the facts that \textit{(i)} it is possible to find examples where the change in the galactic properties can reduce $\fesc$ significantly, and \textit{(ii)} that such a change is accompanied by a change in \Lya spectral properties. 
The questions now are if these changes in $\fesc$ are sufficient to reproduce the observed sudden drop in \lya\ fraction $X_{\rm Ly\alpha}$ without an evolving IGM, and if -- or rather -- in which parameter range these changes in galactic properties are realistic.

\subsection{Escape fraction variation consistent with the change in $X_{\rm Ly\alpha}$}
As can already be seen from the previous section and Fig.~\ref{fig:fesc_NHI_VEL}, the measured drop in $X_{\rm Ly\alpha}$ of $\sim 50\%$ ($\sim 70\%$) for $z\sim 6\rightarrow 7$ ($z\sim 6\rightarrow 8$) can be reproduced entirely by a change in galactic properties. To explore systematically for which parameters this is the case, we ran a grid of models and show explicitly in Figure~\ref{fig:cont_NHI_VEL} the change in escape fraction that is produced by a change in column density  \NHI\ and outflow velocity \vexp\,. We furthermore draw black contour lines representing the change needed to reproduce the observed $X_{\rm Ly\alpha}$ drop for a fixed choice of dust ($\tau_{\rm d}=0.35$). 

In the left panel of Fig.~\ref{fig:cont_NHI_VEL}, we vary the column density at a fixed outflow velocity of 50 ${\rm km/s}$. Starting at low column density e.g. \NHI$= 10^{17-18}\, {\rm cm^{-2}}$ at $z=6$, a
rather large increase by $\sim  1-2$ order of magnitudes is required to obtain the required $\sim$ halving of $\fesc$ by $z=7$. On the other hand, for larger column densities $\log_{10}N_{\rm HI}/{\rm cm}^{-2}> 19$ a significantly smaller increase of $< 1\,$dex in neutral hydrogen column density is necessary, sometimes $\lesssim 0.1\,$dex.

Similarly, in the right panel of Fig.~\ref{fig:cont_NHI_VEL} we show what change in $\fesc$ a change in outflow velocity produces while fixing \NHI$= 10^{21}\, {\rm cm^{-2}}$.
We see here contours representing the observed drop in $X_{\rm Ly\alpha}$ are located in the lower right part, which is opposite to the left panel. This shows that in order to mimic reionization with only the outflow velocity, higher outflows are required at low redshift ($z=6$). The required change in outflows is about $\leq 100 {\rm km/s}$ between these redshifts, which is somewhat moderate. While these models have fixed dust, outflow (left) and column density (right), we can easily predict the corresponding change in these contours for different choice of parameters. For instance, $\fesc$ increases at lower dust values, which means that the contours in the left and right panel would be shifted accordingly to upper and lower part of the grid. Higher outflows increase $\fesc$ and hence the contours would shift to the lower part in the left panel, whereas higher column density would shift the contours in the right panel to the upper part.

In this section, we showed that both a change in column density and a change in outflow velocity can reproduce the observed drop in \Lya detections at $z\gtrsim 6$ usually attributed to an increased neutral hydrogen content of the IGM at these epochs.
We furthermore explored how the change in the galactic properties can mimic an increasing neutral fraction with a change in $\fesc$. A change consistent with the measured drop of the \Lya fraction towards at $z>6$ can be achieved either by boosting the column density by $\sim 0.1$ dex for $N_{\rm HI}\gtrsim 10^{19}\,$or suppressing outflows towards high redshift. We next discriminate between these scenarios using the change in the spectral properties.

\subsection{\lya\ spectral line properties variation as a function of the galactic properties}
As already visible from Fig.~\ref{fig:fesc_NHI_VEL}, the change in the galactic properties that mimics reionization doesn't only produce a different $\fesc$, but also changes significantly the \lya\ spectral line properties which can be tested against observations.
\begin{figure*}[ht!]
    \centering
    \includegraphics[width=0.95\textwidth]{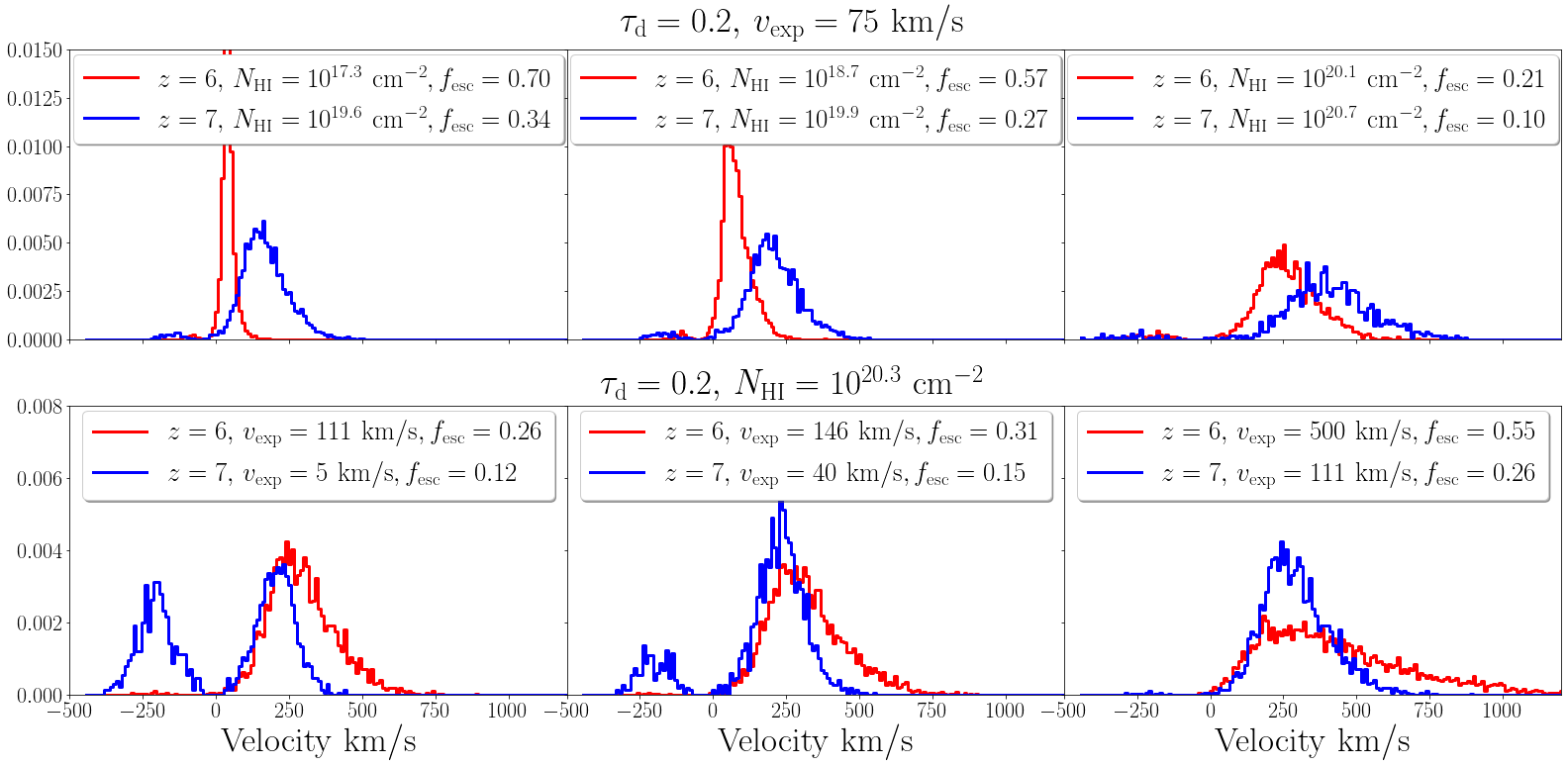}
    \caption{Several examples for the change in spectral properties between $z=6$ (red) and $z=7$ (blue) as the galactic properties change at a fixed amount of dust $\tau_{\rm d}=0.2$ with a pixel resolution of $10\, {\rm km/s}$. Top row shows the change in column density \NHI\ at fixed outflows \vexp\ whereas the bottom row shows the opposite as quoted in the legend and subtitles. In all panels, the difference in $\fesc$ is equivalent to the observed drop in  \lya\ fraction $X_{\rm Ly\alpha}$ between $z=6$ and $z=7$. In the top row, the line width and offset both increase as the column density increases towards high-z, whereas the bottom row indicates that the line width and offset both decrease as the velocity decreases towards high-z. Note that we focus on the most extreme scenario in which the visibility on \Lya at $z>6$ is purely due to galaxy evolution. Thus, an unobscured IGM transmission was assumed (see \S\ref{sec:conclusion} for a discussion of this effect).}
    \label{fig:specs_ex}
\end{figure*}

Figure~\ref{fig:specs_ex} shows several examples for the variation in \lya\ spectral line properties between redshift $z=6$ and $z=7$ at same amount of dust, $\tau_{\rm d}=0.2$. Top row shows the resulting spectral changes due to changing the column density \NHI\ at fixed outflows \vexp\ whereas the bottom row depicts the opposite scenario.  All these examples possess the required $\sim \times 1/2$ $\fesc$ difference between $z=6$ and $z=7$ to mimic the drop in the observed \lya\ fraction $X_{\rm Ly\alpha}$ as explained in the previous section.

In the top panels of Fig.~\ref{fig:specs_ex}, we show that the scenario of changing the column density to mimic reionization suggests that the line width and offset both increase towards high redshift. It is also noted that the line is broader at higher column densities. Interestingly, the second scenario of changing the outflows shown in the lower row of Fig.~\ref{fig:specs_ex} indicates exactly the opposite that broader lines exist at higher outflow velocities, i.e., at lower redshifts.
This is due to the fact that increasing the column density, increases also the escape frequency offset at which \Lya can escape through excursion \citep{Neufeld1990}. On the other hand, for outflowing material a fraction of photons are `backscattered' and obtain $\sim 2$ the expansion velocity \citep{Verhamme2006A&A...460..397V,Dijkstra2006}. This produces the extended red wings in Fig.~\ref{fig:specs_ex} and implies that a lower outflow velocity produces narrower line widths.
%This is due to the fact that $\fesc$ increases with increasing outflows, and reduction in the outflows is required to mimic reionization. 
It is also evident that the blue peak appears more prominently with decreasing outflows to small values, consistent with previous studies \citep{Bonilha1979}.  This means that the two considered scenarios (changing \NHI\ or \vexp\,) to mimic reionization can be distinguished with observations of \Lya spectra. We explore this in detail in the next section.
\begin{figure*}[ht!]
    \centering
    \includegraphics[scale=0.4]{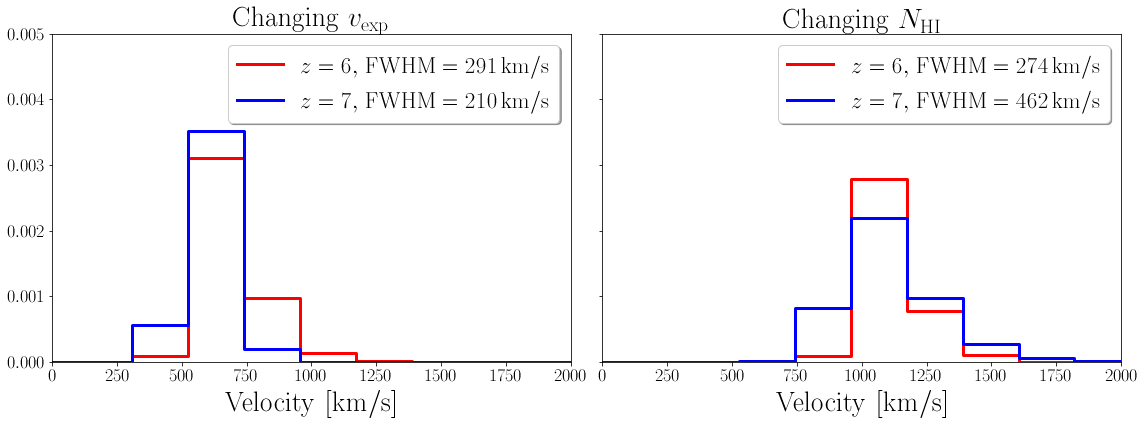}
    \caption{Example for stacks of randomly selected 52 mock spectra at $z=6$ (red) and 19 mock spectra at $z=7$ (blue) with $\fesc(z=7) \approx 0.5\fesc(z=6)$ in agreement with the measured drop of $X_{\rm Ly\alpha}$.  
    We assembled the stacks following the method outlined in~\citet{Pentericci:2018} with spectral resolution $R=1390$ (corresponding to $\Delta=216\, {\rm km/s}$). All these spectra are either obtained by changing only outflows \vexp\ while keeping the column density \NHI\ fixed (left) or changing only the column density \NHI\ while keeping outflows \vexp\  fixed. In all cases, dust is kept fixed. In either scenario, the FWHM of stacks at z=6 is consistent with values reported in~\citet[FWHM($z \sim 6$) = 300$\pm30$]{Pentericci:2018} . The FWHM of stacks obtained by changing outflows \vexp\ at z=7 is also consistent with~\citet{Pentericci:2018} measurements (FWHM($z \sim 7$) = 220$\pm25)$  while those produced by changing the column density \NHI\ are not, and hence indicating that changing the outflow \vexp\ can naturally lead to a higher \Lya escape as well as broader lines at lower redshifts as observations indicate. }
    \label{fig:stacks}
\end{figure*}
\\
\subsection{Comparison to observations}
\label{sec:comparison_obs}
We now use observations to discriminate between our two scenarios. Using the CANDELSz7 survey, \citet{Pentericci:2018} have measured the line widths of two stacks of 52 sources with  $\langle z 
\rangle \sim 6$ and 19 other sources with $\langle z \rangle \sim7$, and found that their full width at half maximum ($\rm FWHM$) are equal to $300\pm 30 \,{\rm km/s}$ and $220\pm 25 \,{\rm km/s}$, respectively. This shows that the\textbf{se} observations indicate that line width is approximately constant or slightly decreasing with increasing redshift. 
Recalling the results of the previous section, this automatically rules out the scenario of changing the column density \NHI\ to mimic reionization since it predicts that the line width increases towards high redshift. 
%In the remainder of this section,

We now explore this more quantitatively as well as check whether a change in outflow velocity or the column density and an associated drop in escape fraction is consistent with the line width measurements of \citet{Pentericci:2018}.

To do so, we attempt to follow the recipe presented in~\citet{Pentericci:2018} to produce stacks of mock spectra  using our model at $z=6$ and $z=7$. Using an initial number of photons of $N_{\rm photon}=10^{4}$, we run grid of 1,125 models over 5 different amount of dust ($\tau_{\rm d} = 0.1-0.5$), 15 outflows (\vexp\ $= 5-500 \, {\rm km/s}$) and 15 column density ($\log_{10} N_{\rm HI}= 17-21\, {\rm cm^{-2}}$); all equidistantly spaced. By considering all possible combinations at fixed column density and dust content, we then select those whose difference in $\fesc$ is equivalent to the observed drop in $X_{\rm Ly\alpha}$ between $z=6$ and $z=7$. We find 650 or 1036 combination of models at $z=6$ and $z=7$ satisfying these requirements in changing the outflows or column density scenarios, respectively. Out of these models, we ignore combinations that are inconsistent (i.e. with FWHM $>\pm 2\sigma$) with~\citet{Pentericci:2018} measurements at $z\sim6$. This reduces the number of models combination to 143 and 201 in the case of changing the outflows and column density, respectively. This means that the presented results here by construction consistent with $z=6$ observations, and we aim to explore the different scenarios predictions at high-z as compared with measurements.   Similar to \citet{Pentericci:2018}, we bin all spectra using a spectral resolution of $R=1390$, corresponding to velocity resolution of $\Delta {\rm v}=216\, {\rm km/s}$, and produce median stacks of randomly selected 52 spectra at $z=6$ and only 19 spectra at $z=7$. We also add noise to each individual spectrum drawn from Gaussian distribution of zero mean and standard deviation set by the  signal-to-noise ratio of $S/N=5$ per pixel (i.e., $\sigma_i = F_i / 5$ where $F_i$ is the flux of pixel $i$). We show two random examples for stacks of this procedure at $z=6$ (red) and $z=7$ (blue) in Figure~\ref{fig:stacks} as obtained by changing only the outflows (left) or changing only the column density (right).
While both scenarios produce stacks that have consistent FWHM values with observations at $z=6$, the changing outflows scenario yields also a consistent FWHM with $z=7$ measurements. This also confirms that changing the column density predict very high FWHM $>$ 400 $km/s$ at $z=7$. As quoted in the legend for changing the outflows scenario (left), $z=6$ lines are broader than those at $z=7$, and consistent with~\citet{Pentericci:2018}. The lines are highly asymmetric with extended red wings. The blue peaks disappear due to the poor resolution.

\begin{figure}[h!]
    \centering
    \includegraphics[scale=0.4]{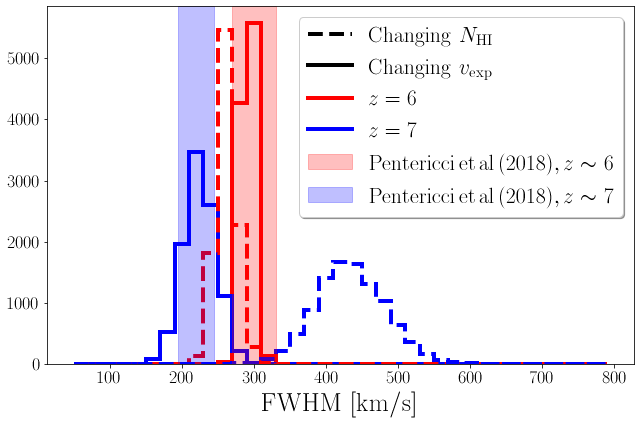}
    \caption{Line width distribution using randomly generated 10,000 stacks at $z=6$ (red) and $z=7$ (blue) obtained by changing only the outflows (solid) or changing only the column density (dashed). Shaded red and blue areas show~\citet{Pentericci:2018} measurements at $z\sim 6$ and $z\sim 7$, respectively. It is evident that the scenario of changing the outflows is consistent with the observations, while changing the column density scenario predict very high FWHM values at $z\sim 7$.}
    \label{fig:width}
\end{figure}

To quantify the width using the above recipe, we generate randomly 10,000 combination of stacks at $z=6$ and $z=7$ from the total number of model combinations (i.e. 143 and 201) in each scenario and compute their widths. We show the resulting width distribution at these redshifts in Figure~\ref{fig:width}. Results from changing outflows  and changing column density scenarios are presented by dashed and solid lines. Shaded red and blue areas show~\citet{Pentericci:2018} measurements at $z\sim 6$ and $z\sim 7$, respectively. From this exercises alone, we constrain the FWHM, using changing column density scenario, to $261.7\pm 14.3$ km/s at $z=6$ and $434.9 \pm 47.8$ km/s at  $z=7$,  and using changing outflows scenario, to $291.5 \pm 7.9$  km/s  $z=6$  and $224.6 \pm 22.8$  km/s   at $z=7$. This evident that, over all possible combinations and the prior range assumed, the changing outflows scenario produces consistent width distribution within the 1-$\sigma$ level of~\citet{Pentericci:2018} measurements. This confirms that changing the column density predicts inconsistent width distributions. It might be noteworthy that while we use the \citet{Pentericci:2018} results in this section, other studies of the spectral properties at that redshift might obtain different results due to a different sample selection.

\begin{figure}[h!]
    \centering
    \includegraphics[scale=0.4]{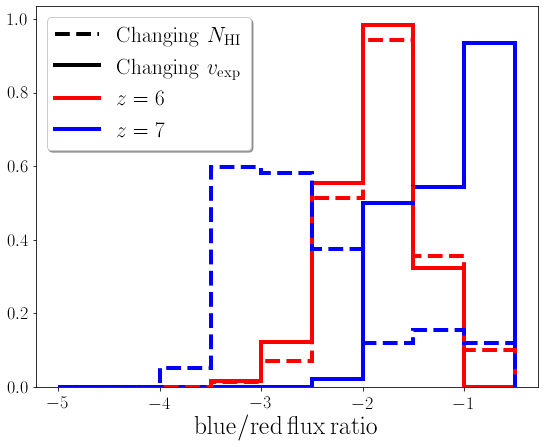}
    \caption{Predictions for the blue/red flux ratio at $z=6$ (red) and $z=7$ (blue) for changing the column density scenario (dashed) or changing the outflows (solid). Both scenarios produce similar flux ratio at $z=6$. Changing the outflows scenario predicts higher flux ratio at $z=7$, indicating the presence of more blue peaks at high redshift, which is opposite to the change in column density scenario. }
    \label{fig:BRFR}
\end{figure}

We now use these scenarios to make predictions for the blue/red flux ratio, which is defined as the total blue line flux divided by the total red flux. We perform these predictions at the level of individual spectra not with stacks, since the blue peaks disappear due to the poor resolution. We show the blue/red flux ratio distribution from total number of model combinations in Figure~\ref{fig:BRFR}. Both scenarios produce similar flux ratio at $z=6$. However, changing the outflows scenario predicts higher flux ratio at $z=7$, indicating the presence of more blue peaks at high redshift, which is opposite to the change in column density scenario. Note that for this paper we explore the somewhat extreme scenario where the entire change of the \Lya visibility is due to galaxy evolution alone, i.e., we assume a completely transparent IGM. In reality, the majority of the IGM at $z\gtrsim 5$ is already opaque to \Lya photons blueward of line-center. This means that an enhancement of the blue peak towards higher $z$ can be translated to a lower observed \Lya visibility (even with non-evolving IGM) and, thus, to a reduction of the required change of galactic properties in order to reproduce the observed change in EW(\lya). We will discuss this further in \S\ref{sec:conclusion}.

\section{Concluding remarks}
\label{sec:conclusion}
The decrease in \Lya visibility for $z>6$ is commonly interpreted as a change in IGM opacity due to the Epoch of Reionization.
As an alternative, we have explored scenarios of how the change in the galactic properties can naturally lead to drop in the photon escape fraction that is equivalent to the observed drop in \lya\ fraction while keeping the IGM fixed at $z\geq6$. 
We have considered two scenarios: changing the column density \NHI\, or the outflow velocity \vexp. We found that decreasing the column density by only $\lesssim 1$\,dex ($\lesssim 0.1\,$dex for $N_{\rm HI}\gtrsim 10^{19}\,{\rm cm}^{-2}$) or increasing the outflow velocity by $\sim 100\kms$ with decreasing redshift can both successfully reproduce the observed drop in the \Lya fraction, and thus, `mimic' an increasing IGM neutral fraction.
Note that these exact values depend on the observed drop in $X_{\rm Ly\alpha}$. We have adopted values consistent with most studies (cf. \S\ref{sec:methods}) but note that  recent work by~\citet{Kusakabe:2020} have reported much lower value of \lya\ fraction at $z\sim6$ of $X_{Ly\alpha}=0.13$, which is less than~\citet{Stark:2011} by a factor of 4. Naturally this would lead to a smaller evolution in the explored galactic properties but since the primary goal of this study is to show whether in principle an observed drop in \Lya visibility can be explained by galaxy evolution (independent of the exact redshift it occurs at) these yet existing observational differences do not alter our conclusion.

To differentiate between the evolution in galactic and intergalactic properties, we analyze the associated change in \Lya spectral properties.  The line width and offset both increase as the column density and outflows increase. The observed spectral properties can potentially discriminate between these scenarios, which indicate that the broader lines exists in low redshift~\citep{Pentericci:2018}. This automatically rules out the changing column density scenario (cf. \S\ref{sec:comparison_obs}).
On the other hand, the scenario of a change in outflow properties does not alter the spectral width significantly, and is thus compatible with current observations. 
Following \citet{Pentericci:2018}, we generate 10,000 stacks and compare to the observed width results. We find the line width is $291.5 \pm 7.9 \, {\rm km/s}$ at $z=6$ and $224.6 \pm 22.8\, {\rm km/s}$ at $z=7$, which is consistent with the 1-$\sigma$ level of the \citet{Pentericci:2018} measurements. 
We predict that such a scenario of a change in outflow properties implies a larger flux on the blue side of \Lya towards higher redshifts.
While in principle this could be directly detectable -- and there has been an increasing number of blue peaks at high-$z$ has been detected \citep{Hu2016ApJ...825L...7H,Songaila2018ApJ...859...91S,Matthee2018A&A...619A.136M,Bosman2020ApJ...896...49B} -- the IGM already at $z\lesssim 5$ is already mostly opaque to \Lya photons on the blue side altering systematically the observed spectra \citep{Laursen2011ApJ...728...52L,Hayes2020,Byrohl2020}.
Such an evolution would thus have to be indirect, i.e., through the (change in) \Lya halo properties.

Throughout this study, we model the complex radiative transfer through the galactic and circumgalactic medium by the simple concentric, outflowing shell. While this `shell-model' has been shown to reproduce observed \Lya spectra well\footnote{Something that cannot be claimed when using more complex input geometry, e.g., from galactic simulations \citep[see discussion in][]{2018ApJ...862L...7G,Mitchell2020} which further justifies the usage of the `shell-model' in this study.}, there is an ongoing discussion in the literature why this is, and what the shell-model parameters represent \citep[e.g.,][]{Gronke2017,Orlitova2018A&A...616A..60O,Li2020}. 
While a full discussion of the problematic is beyond the scope of this work, we want highlight some points most relevant for this study. 
In particular, both theoretical \citep{Dijkstra2016ApJ...823...74D,Eide2018ApJ...856..156E,KakiichiGronke2020} as well as observational work (Vielfaure et al., in prep.) points towards the fact that \Lya spectra are in a way an extremum statistics, i.e., they are heavily weighted towards low opacity channels. This is maybe unsurprising as \Lya photons most easily escape through these `pathways of least resistance'. Specifically, from this theoretical work it became clear that \Lya spectra can be shaped by the lowest column densities channels -- even when this is not along the line-of-sight towards the observer \citep[e.g.,][]{Eide2018ApJ...856..156E}\footnote{While observational confirmation is difficult as it requires an independent tracer of the HI column density, for instance, GRB afterglow spectra offer an attractive opportunity \citep{Vielfaure2020A&A...641A..30V}.}.
This implies for our study that not the average galactic properties but instead the `extreme' (in terms of opacity) has to evolve in order to mimic reionization. As such lower density or higher velocity channels can occur on relatively short timescales ($\lesssim$ tens of Myr), for instance, due to a burst of star formation \citep{1989ApJ...345..372N,2017MNRAS.466...88S,2019ApJ...884..133F}, the required change in the `shell model' parameters -- in particular the outflow velocity -- from $z\sim 6$ to $\sim 7,\,8$ ($> 200\,$Myr) is in fact not unlikely.
We expect future studies targeting the connection of the `shell-model' to more realistic gas geometries to allow improved estimates on their variability. 
While we argue that the `shell-model' represents a good model for \Lya radiative transfer in order to reproduce observed spectra, this is clearly not the case for the surface brightness (SB) profile \citep[e.g.,][]{2020ApJ...901...41S}. A realistic SB profile could lead to additional non-observed \Lya photons due to the SB limit of observations, however, this effect is beyond the scope of this paper, i.e., here we assume a non-evolving SB profile.

A limitation of this study is the homogeneous treatment of the galaxy population.  A more realistic approach is to develop a novel approach, such abundance matching, to link the equivalent width distribution to the galaxy mass or stellar mass function at these epoch, and then study what change is required in the whole galaxy population to mimic the whole equivalent width distribution. 
We leave exploring possibilities to undertake such approach in future works. 

Additionally, the dust has been also assumed fixed in our analysis. To first order expected evolution of dust optical depth for \Lya photons is to follow the metallicity, i.e., to decrease towards high redshift. This in turn would increases the $f_{\rm esc, Ly\alpha}$, and goes in the opposite direction to the observed drop in $X_{\rm Ly\alpha}$.
In our both scenarios, it is still possible to combine the increase of dust towards high redshift with the increase of column density of decrease in outflows, but these changes in the galactic properties would be larger to offset the evolution in dust. Given the uncertainty in dust and metalicity at these high redshift epochs, and general uncertainty of how does interacts with \Lya photons\footnote{How susceptible \Lya photons are to dust depends heavily its distribution and is focus of a large body of literature \citep[e.g.,][]{Neufeld1991,Hansen2006MNRAS.367..979H}.}, we have compared models at the same level of dust optical depth.\\
% and metallicity. 

In summary, we have shown that the change in \Lya visibility towards higher redshift can be attributed to a change in galactic properties and do not require a change in IGM properties. Specifically, we find that both a modest decrease in column density \citep[in agreement with][]{Sadoun:2017} or an increase in outflow velocity in the galaxies' evolution can lead to a increase in \Lya escape fraction, and thus, to the observed drop in detected \Lya emission towards at $z>6$. 
We furthermore found that this degeneracy between IGM transmission and galactic \Lya radiative transfer can be broken using the emergent \Lya spectral properties. In particular, we found that the scenario of a change in column density leads to unnatural wide \Lya profiles and is ruled out by existing data -- but that the change is outflow properties is not.
We predict that in such a case there will be more blue flux emergent from galaxies towards redshifts which can be detectable either directly or through an evolution in \Lya halo properties.

Naturally, a fast evolution of the IGM neutral fraction at $z>6$ is expected and observed using other, independent probes, we caution that the uncertainty of galactic \Lya radiative transfer should be taken into account when constraining this evolution using the observed \Lya equivalent width distribution or luminosity function.

\section*{Acknowledgements} 
The authors acknowledge helpful discussions with Charlotte Mason, Daniel Stark, Edmund Christian Herenz and Steven Finkelstein, and thank the referee for constructive comments.
This research made use of matplotlib \citep{Hunter:2007}, SciPy \citep{Virtanen_2020}, IPython \citep{PER-GRA:2007}, and NumPy \citep{harris2020array}. Part of this work was performed during a research visit to the Space Telescope Science Institute (STScI) where support was provided by the STScI Director’s Discretionary Fund.
Simulations and analysis were performed at NMSU's DISCOVERY supercomputers.  This work also used the Extreme Science and Engineering Discovery Environment (XSEDE), which is supported by National Science Foundation grant number ACI-1548562, and computational resources (Bridges) provided through the allocation AST190003P. MG was supported by NASA through the NASA Hubble Fellowship grant HST-HF2-51409 and acknowledges support from HST grants HST-GO-15643.017-A, HST-AR-15039.003-A, and XSEDE grant TG-AST180036.

\bibliography{sample63}{}
\bibliographystyle{aasjournal}



\end{document}